\documentclass{article}
\usepackage{spconf,amsmath,graphicx,hyperref,xcolor,amsfonts, subcaption,booktabs,placeins}
\usepackage[framemethod=tikz]{mdframed}
\usepackage{xcolor}
\usepackage{soul}

\hypersetup{
    colorlinks=true,
    citecolor=blue,
    linkcolor=blue,
    filecolor=magenta,      
    urlcolor=blue,
}
\begin{document}

\title{VCD: A Video Conferencing Dataset for Video Compression}
\name{
\parbox{\linewidth}{\centering
Babak Naderi, Ross Cutler, Nabakumar Singh Khongbantabam, Yasaman Hosseinkashi, Henrik Turbell, Albert Sadovnikov, Quan Zhou}}

\address{Microsoft Corporation, Redmond, USA} 

\maketitle

\begin{abstract}
Commonly used datasets for evaluating video codecs are all very high quality and not representative of video typically used in video conferencing scenarios. We present the Video Conferencing Dataset (VCD) for evaluating video codecs for real-time communication, the first such dataset focused on video conferencing. VCD includes a wide variety of camera qualities and spatial and temporal information. It includes both desktop and mobile scenarios and two types of video background processing. We report the compression efficiency of H.264, H.265, H.266, and AV1 in low-delay settings on VCD and compare it with the non-video conferencing datasets UVC, MLC-JVC, and HEVC. The results show the source quality and the scenarios have a significant effect on the compression efficiency of all the codecs. VCD enables the evaluation and tuning of codecs for this important scenario. The VCD is publicly available as an open-source dataset at \url{https://github.com/microsoft/VCD}.
\end{abstract}
\vspace{2mm}

\maketitle
\section{Introduction}
Video conferencing has become an essential means of communication, especially with the shift to remote work and learning over the past few years. High-quality video compression is critical for providing a smooth and effective video conferencing experience. Most video codec evaluation relies on benchmark datasets capturing broadcast or cinematic content, which represents video with high spatial resolution, temporal resolution, and visual complexity \cite{bossen_common_2013}. However, video conferencing presents a very different use case from entertainment video. Video conferencing content typically comes from webcams of a range of quality less than studio-quality cameras, with lower spatial resolution temporal resolution, and lower visual complexity compared to traditional video benchmarks. The content in video conferencing is typically a talking person or people in a conference room, typically from stationary cameras, but sometimes from a mobile device.

Despite the key differences between video conferencing and entertainment video, current benchmarking datasets remain focused on the latter.
Datasets such as HEVC Class A-E \cite{bossen_common_2013}, UVG \cite{mercat_uvg_2020}, and MCL-JCV \cite{wang_mcl-jcv_2016} feature 4K and HD video sequences with high spatial and temporal complexity which is useful for assessing codec performance in entertainment applications. However, they do not capture the specific characteristics of real-world video conferencing streams. It is crucial to test video codecs with video conferencing workloads to ensure they are not optimized solely for cinematic content. 

We present the Video Conferencing Dataset (VCD), the first publicly accessible video codec benchmark for video conferencing applications. VCD comprises diverse video sequences originating from mobile and desktop conferencing scenarios, showcasing various spatial and temporal attributes. It encompasses content with and without visual background replacement, a prevalent feature in modern video conferencing setups.
 To establish the utility of VCD, we ran rigorous statistical tests that involved three mainstream datasets and four codecs (H.264/AVC \cite{wiegand_overview_2003}, H.265/HEVC \cite{sullivan_overview_2012}, H.266/VVC \cite{bross_overview_2021}, AV1 \cite{han_technical_2021}). Statistical models reveal interaction effects between dataset choice and video codec performance, emphasizing that conclusions drawn from mainstream datasets may not apply to video conferencing scenarios. 

In Section \ref{sec:related_work}, we discuss related work, and in Section \ref{sec:dataset}, we describe the VCD design. In Section \ref{sec:analysis}, we provide analysis on VCD, and in Section \ref{sec:conclusions}, we provide conclusions.

\section{Related Work}
\label{sec:related_work}

The Joint Video Experts Team (JVET) developed the official test sequences to evaluate H.265/HEVC ~\cite{sullivan_overview_2012} encoder proposals \cite{bossen_common_2013}. It consists of 15 video clips in four classes: B-E. Class B is five 1080p natural video sequences, Class C is four WVGA natural sequences, Class D is four WQVGA natural video sequences, and Class E is three 720p high-quality video sequences of people talking and 4 videos of screen content of various resolutions. 

JVET extended the HEVC test set to support H.266/VVC~\cite{bross_overview_2021} by adding an additional 10 video clips in classes A1, A2, and F \cite{bossen_jvet_2019}. Class A1 and A2 are each four 4K natural video sequences. Class F is two natural sequences and two videos of screen content. 

The Media Communications Lab JND-based Coded Video (MCL-JCV) dataset \cite{wang_mcl-jcv_2016} consists of 30 1080p video sequences. The sequences include 3 genres (cartoon, sports, indoor), 3 semantic classes (people, water, salience), and 3 feature classes (fast motion, camera motion, dark scene). No analysis of spatial or temporal information was used in the design of MCL-JVC, and no description of the camera source or analysis of camera quality was provided. 

The Ultra Video Group (UVG) \cite{mercat_uvg_2020} is a set of 16 4K natural sequences captured at 50 or 120 FPS. It includes spatial and temporal information analysis, which shows the 16 sequences cover a wide range of spatial and temporal information. The dataset was captured with a studio-quality camera.

The Tencent Video Dataset (TVD) \cite{xu_tencent_2021} is a set of 86 4K natural sequences captured with studio-quality cameras. A variety of scenes with static or moving objects are included, but no analysis of spatial or temporal information is done. The Large-Scale Screen Content Dataset \cite{cheng_lscd_2023} provides 714 sequences of screen content, which complements VCD.

\section{Dataset}
\label{sec:dataset}
\begin{figure}
    \centering
  \includegraphics[width = 0.7\columnwidth]{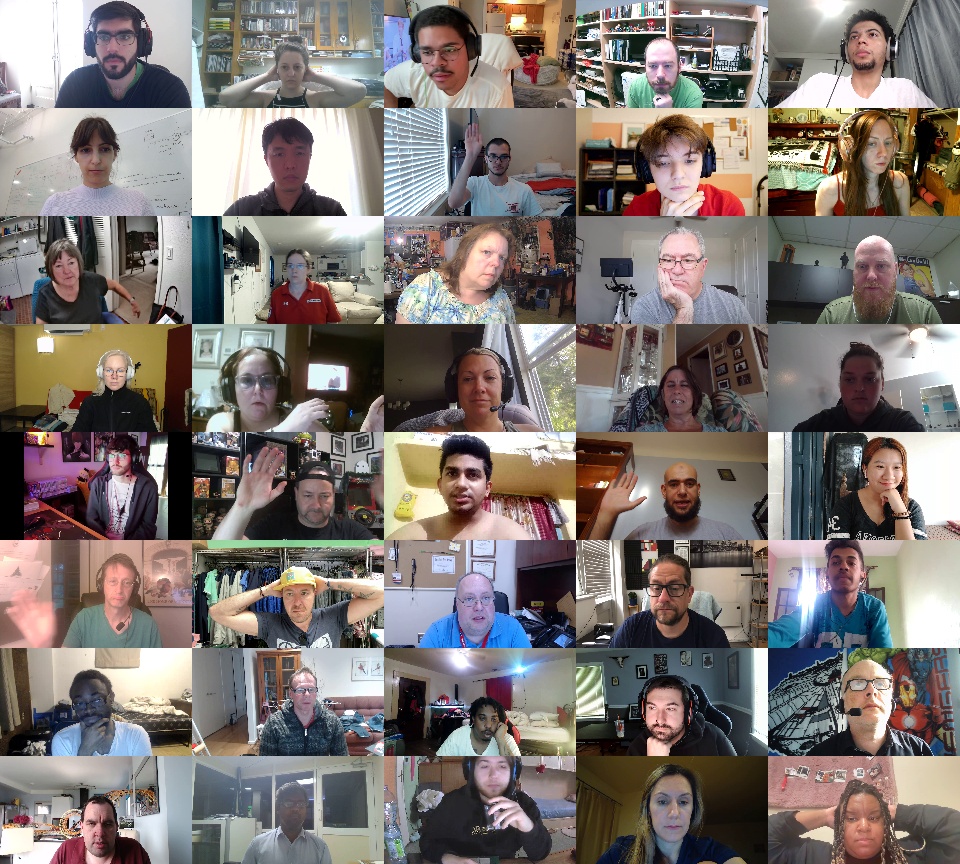}
  \caption{Thumbnail images of 40 sequences belong to Talking Head (TH) without Background effect scenario.}
  \label{fig:s1_thumbnail}
\end{figure}

\begin{figure*}[tb]
    \begin{subfigure}[c]{0.24\textwidth}
        \includegraphics[width=1\textwidth]{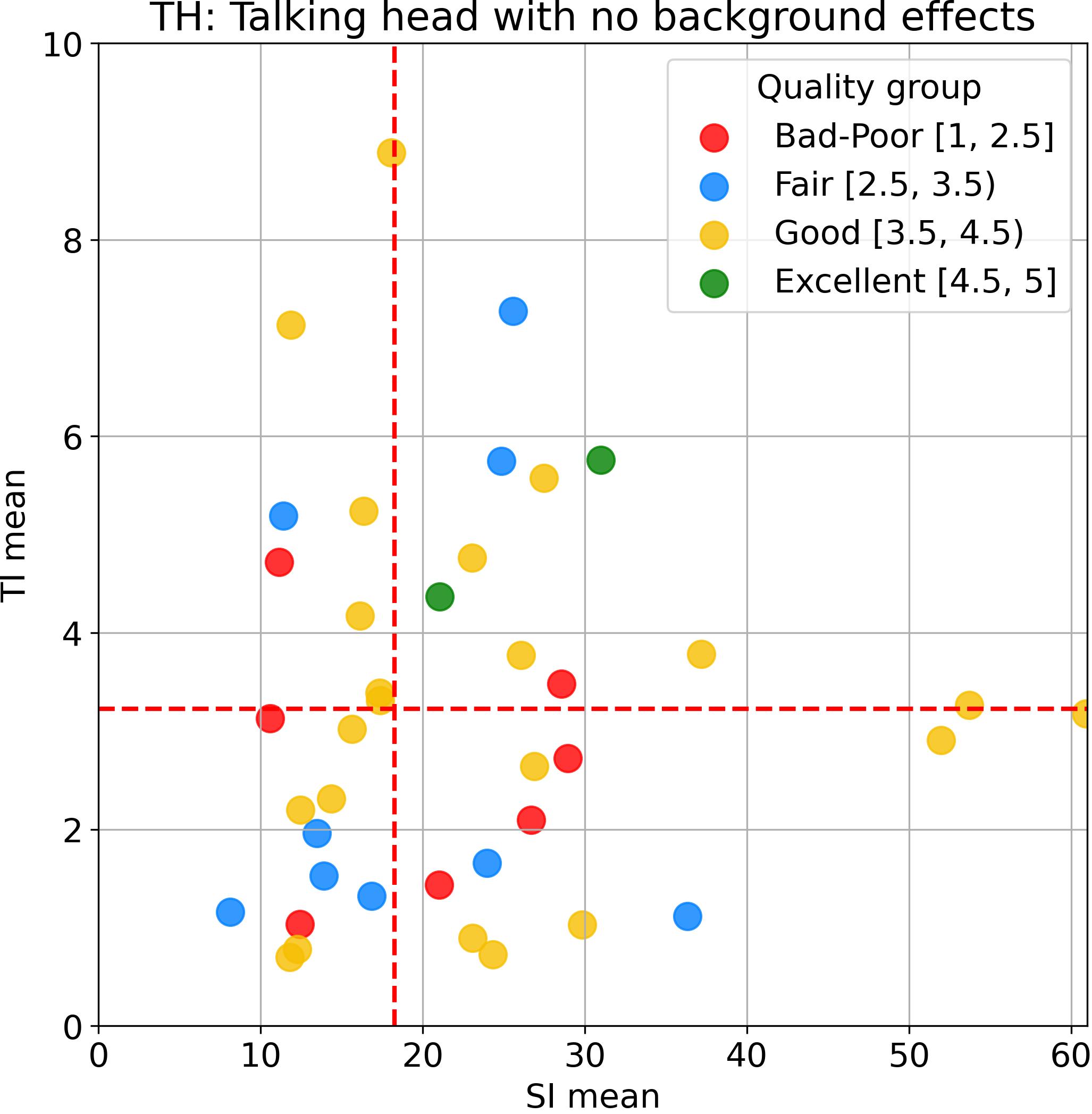}
        \caption{TH}
    \end{subfigure}    
    \begin{subfigure}[c]{0.24\textwidth}
        \includegraphics[width=\textwidth]{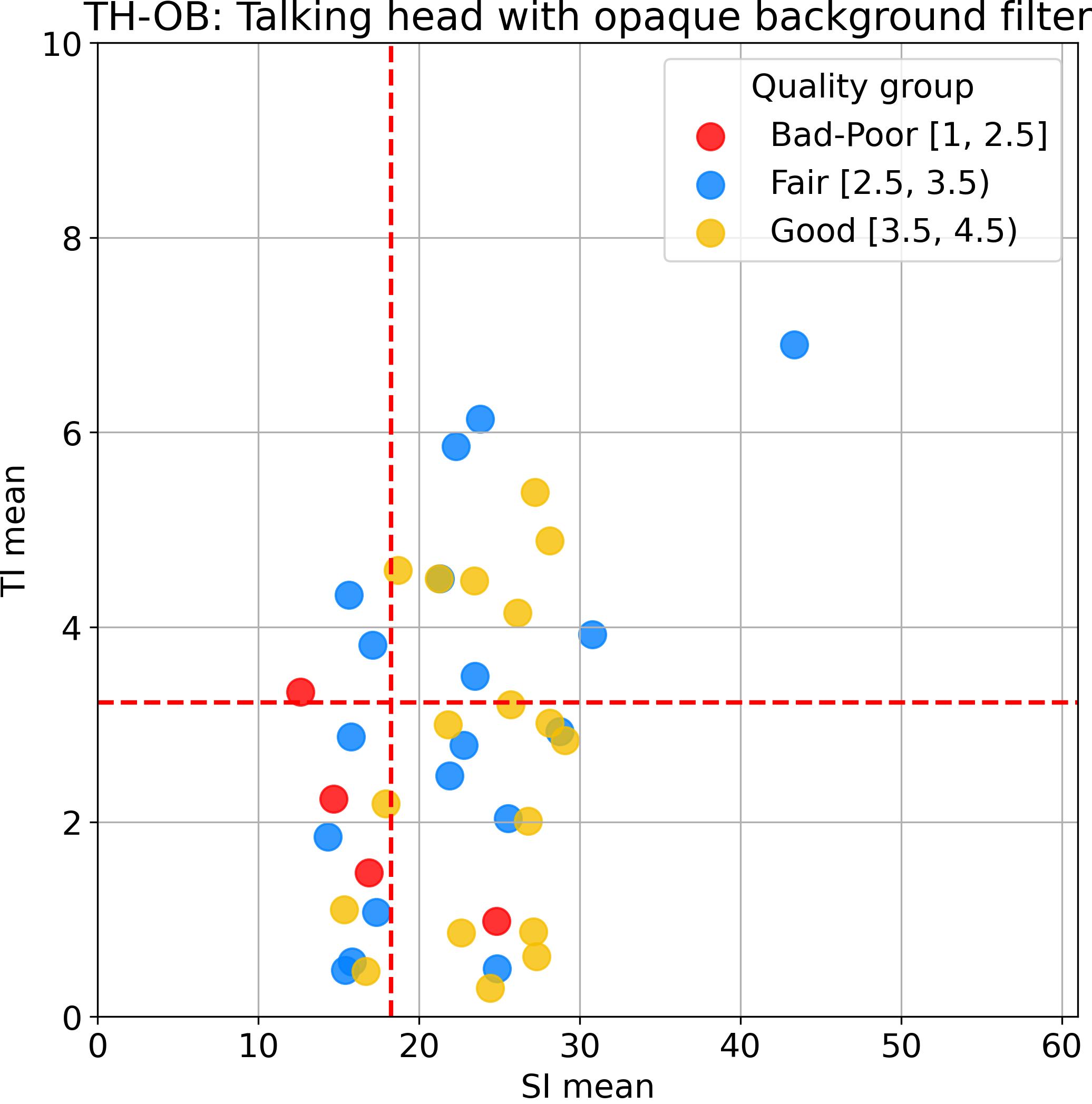}
        \caption{TH-OB}
        \label{fig:thob}
    \end{subfigure}   
    \begin{subfigure}[c]{0.24\textwidth}
        \includegraphics[width=\textwidth]{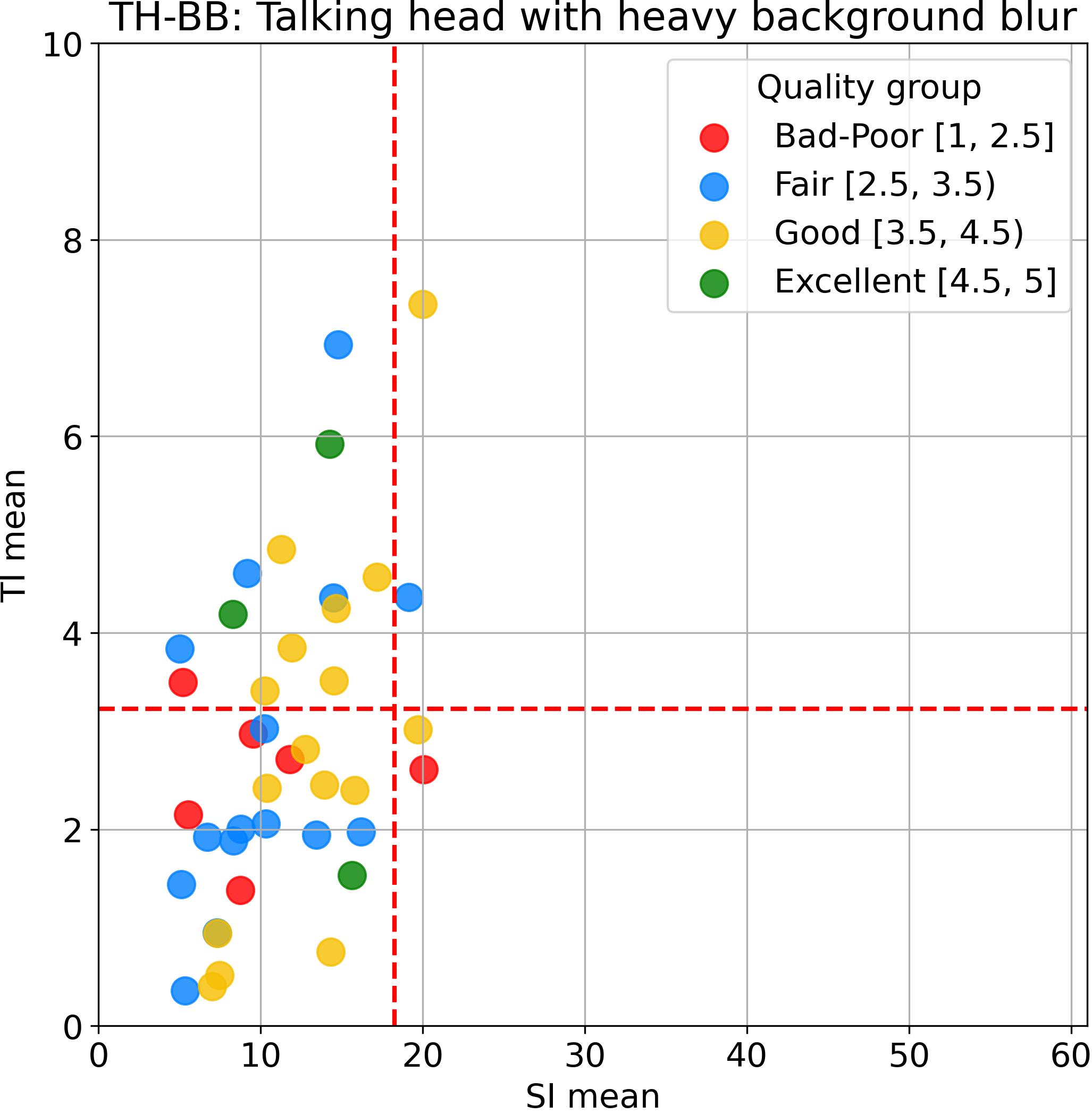}
        \caption{TH-BB}
        \label{fig:thbb}
    \end{subfigure}   
    \begin{subfigure}[c]{0.24\textwidth}
        \includegraphics[width=\textwidth]{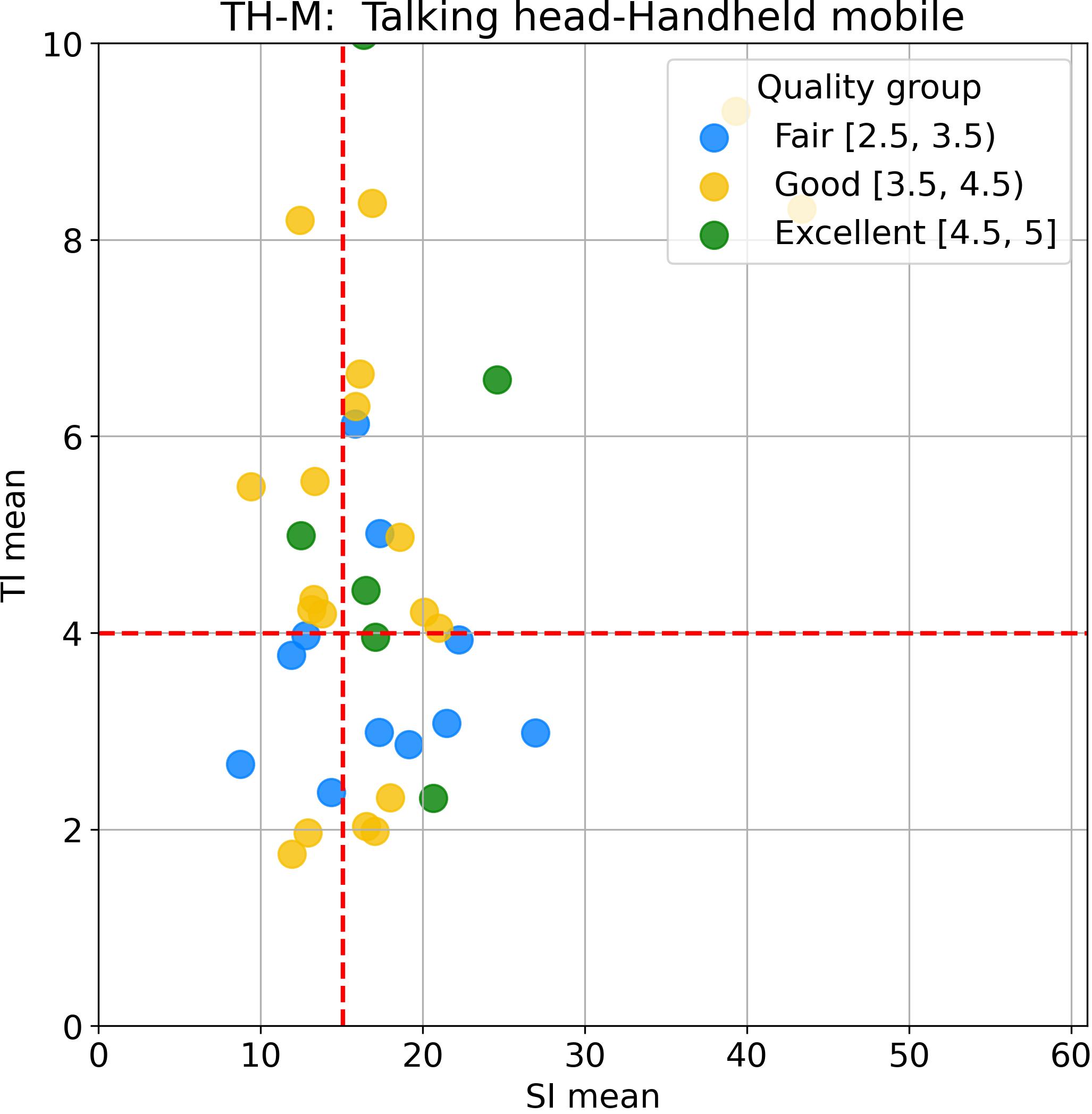}
        \caption{TH-M}
    \end{subfigure}
    \caption{Distribution of Spatial and Temporal information of video sequences in VCD dataset varies by background processing}
    \label{fig:si_ti}
\end{figure*}

\begin{figure}
    \centering
  \includegraphics[width = 0.9\columnwidth]{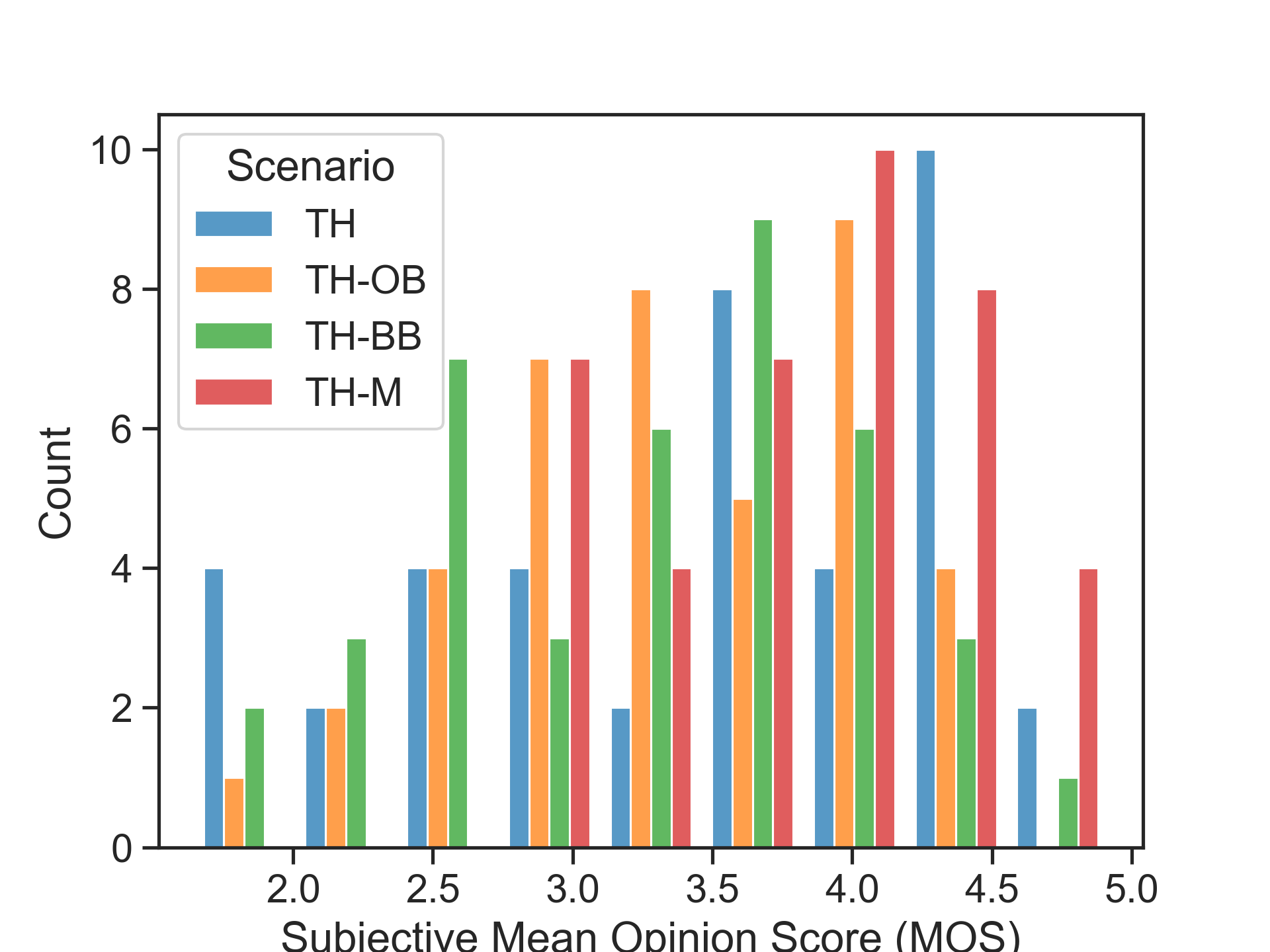}
  \caption{Quality distribution of video sequences in VCD dataset measured in ACR subjective test.}
  \label{fig:quality_distribution}
\end{figure}

VCD consists of 160 talking-head video sequences using mutually exclusive subjects and environments. It is organized in four scenarios, each 40 sequences. The first scenario is Talking Head (TH) videos and includes sequences as they were recorded by each participant's webcam without further processing (see Figure~\ref{fig:s1_thumbnail}). Scenario two refers to Talking Head with Opaque Background filter (TH-OB) where the Microsoft Teams background filter pipeline is applied and the participant's backgrounds are replaced by two popular background images used in Microsoft Teams video calls. Scenario three is similarly processed by Teams' background segmentation pipeline, however with a blurred background filter (TH-BB). The last scenario includes handheld mobile recordings (TH-M) and includes inside and outside recorded videos. At least 30\% of sequences are with active speakers in each scenario. The distribution of temporal and spatial information (TI and SI) \cite{itu-t_recommendation_p910_subjective_2021}\footnote{Values are calculated using https://github.com/VQEG/siti-tools} of video sequences for each scenario are presented in Figure~\ref{fig:si_ti}. In addition, Figure~\ref{fig:quality_distribution} represents the distribution of subjective quality ratings for all sequences in the dataset.

The dataset is released in YUV420p pixel format as 1080p 30 FPS 10-second clips. Besides the sequences, we open-source the subjective Mean Opinion Score (MOS), spatial information (SI), and temporal information (TI) of each sequence.

\begin{figure*}
    \begin{subfigure}[c]{0.245\textwidth}
        \includegraphics[width=1\textwidth]{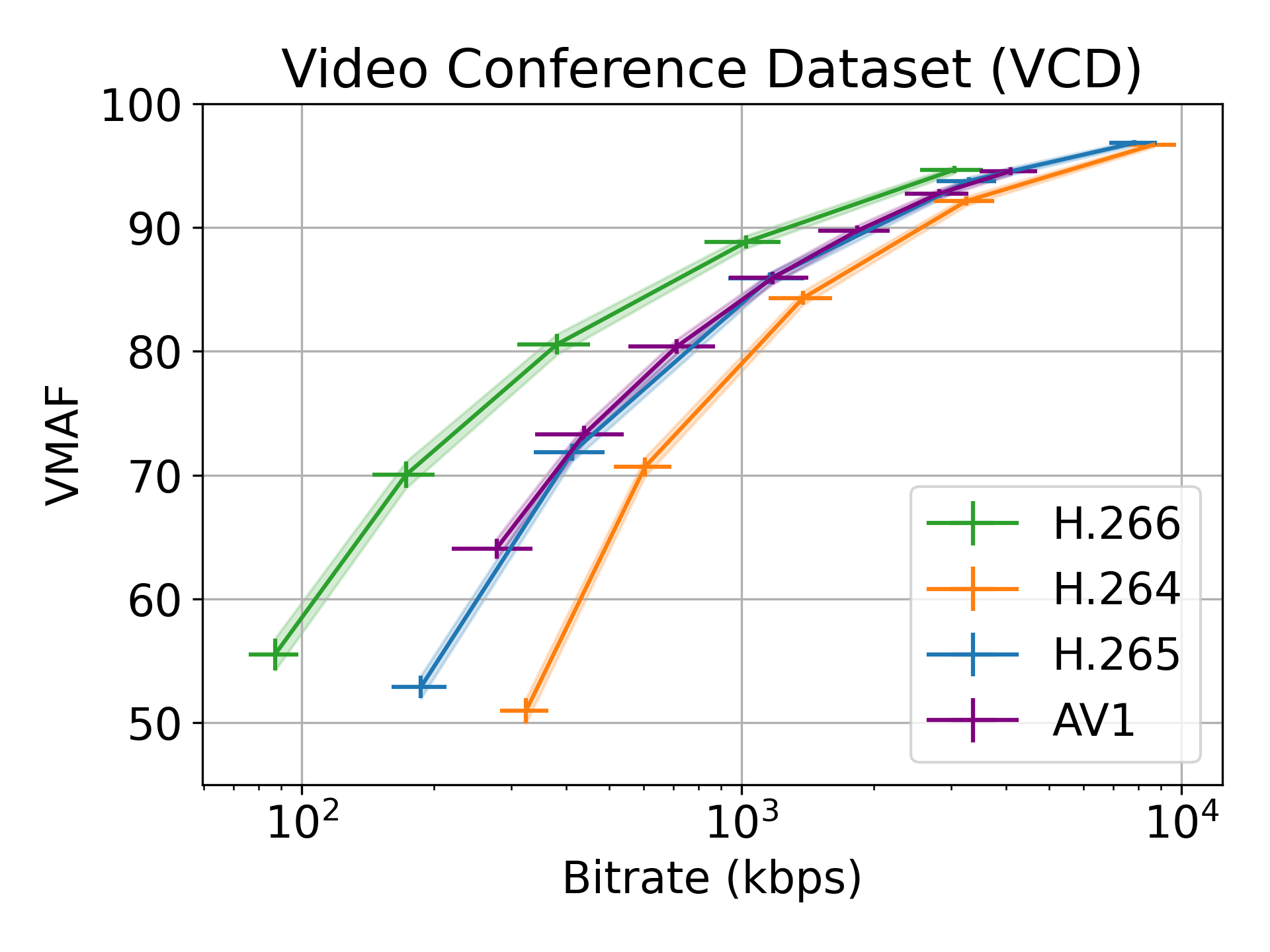}
        \caption{}
        \label{fig:vcd_vmaf}
    \end{subfigure}
    \begin{subfigure}[c]{0.245\textwidth}
        \includegraphics[width=\textwidth]{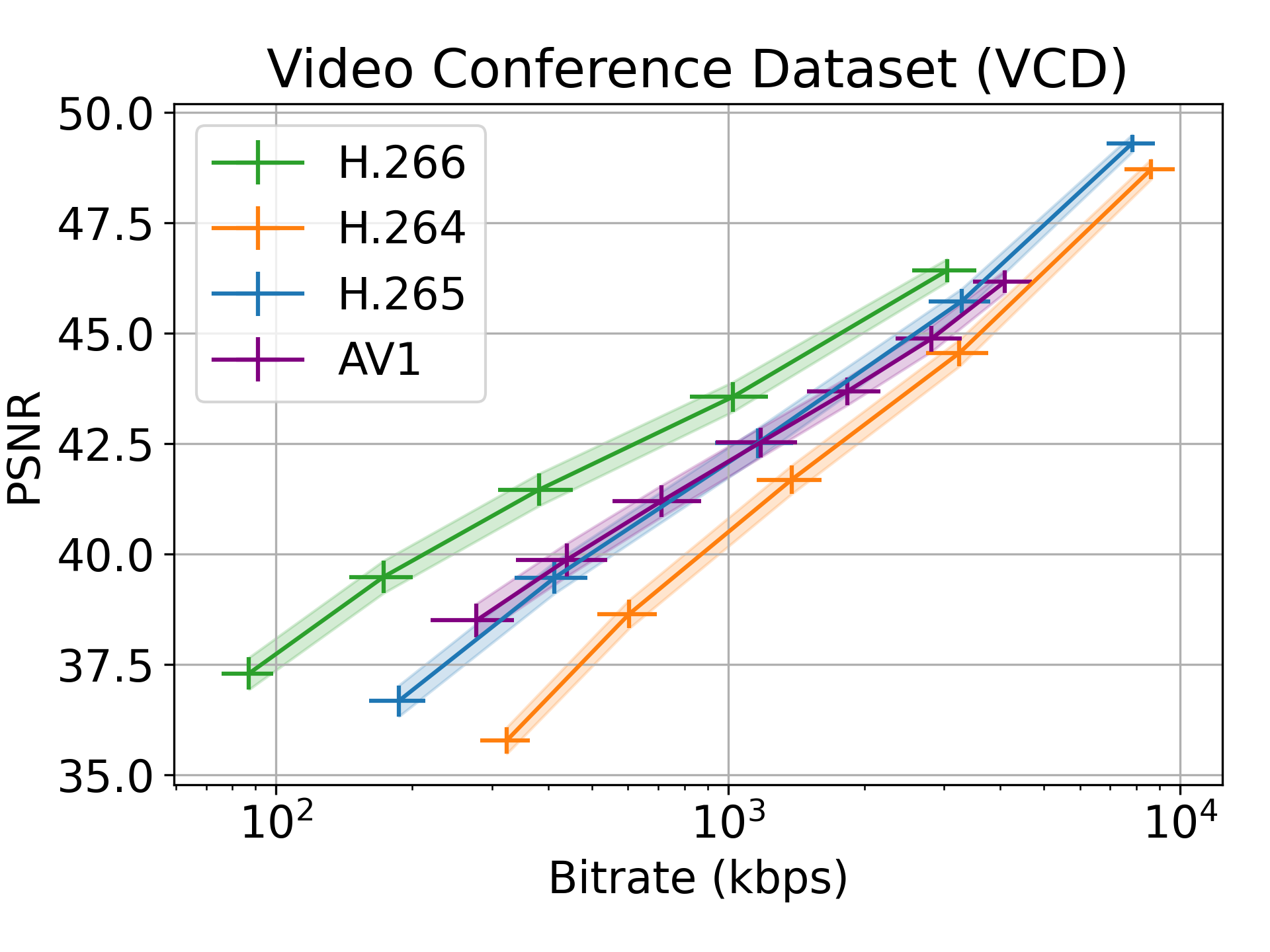}
        \caption{}
        \label{fig:vcd_psnr}
    \end{subfigure}
    \begin{subfigure}[c]{0.245\textwidth}
        \includegraphics[width=\textwidth]{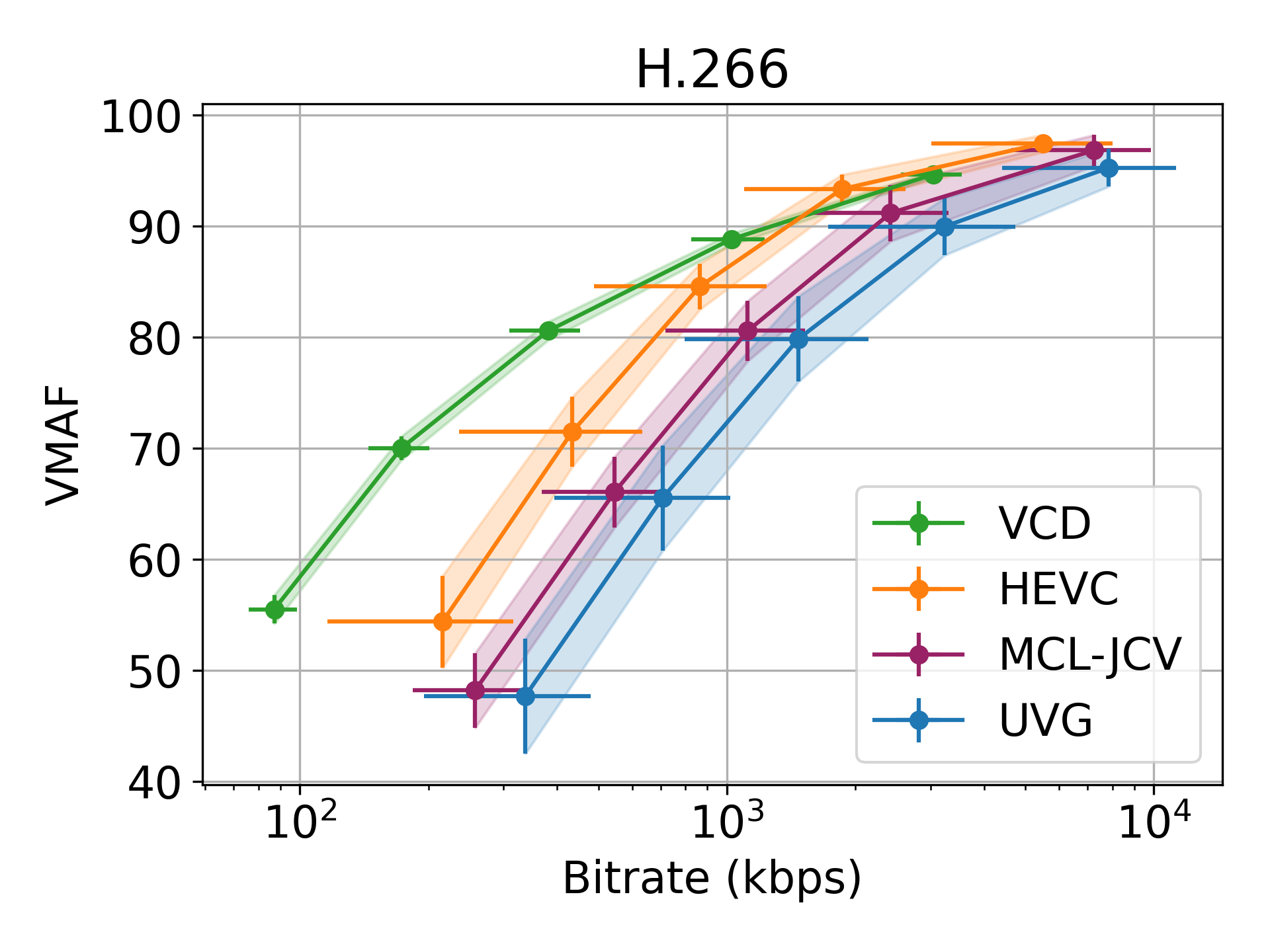}
        \caption{}
        \label{fig:vvc_vmaf}
    \end{subfigure}
    \begin{subfigure}[c]{0.245\textwidth}
        \includegraphics[width=\textwidth]{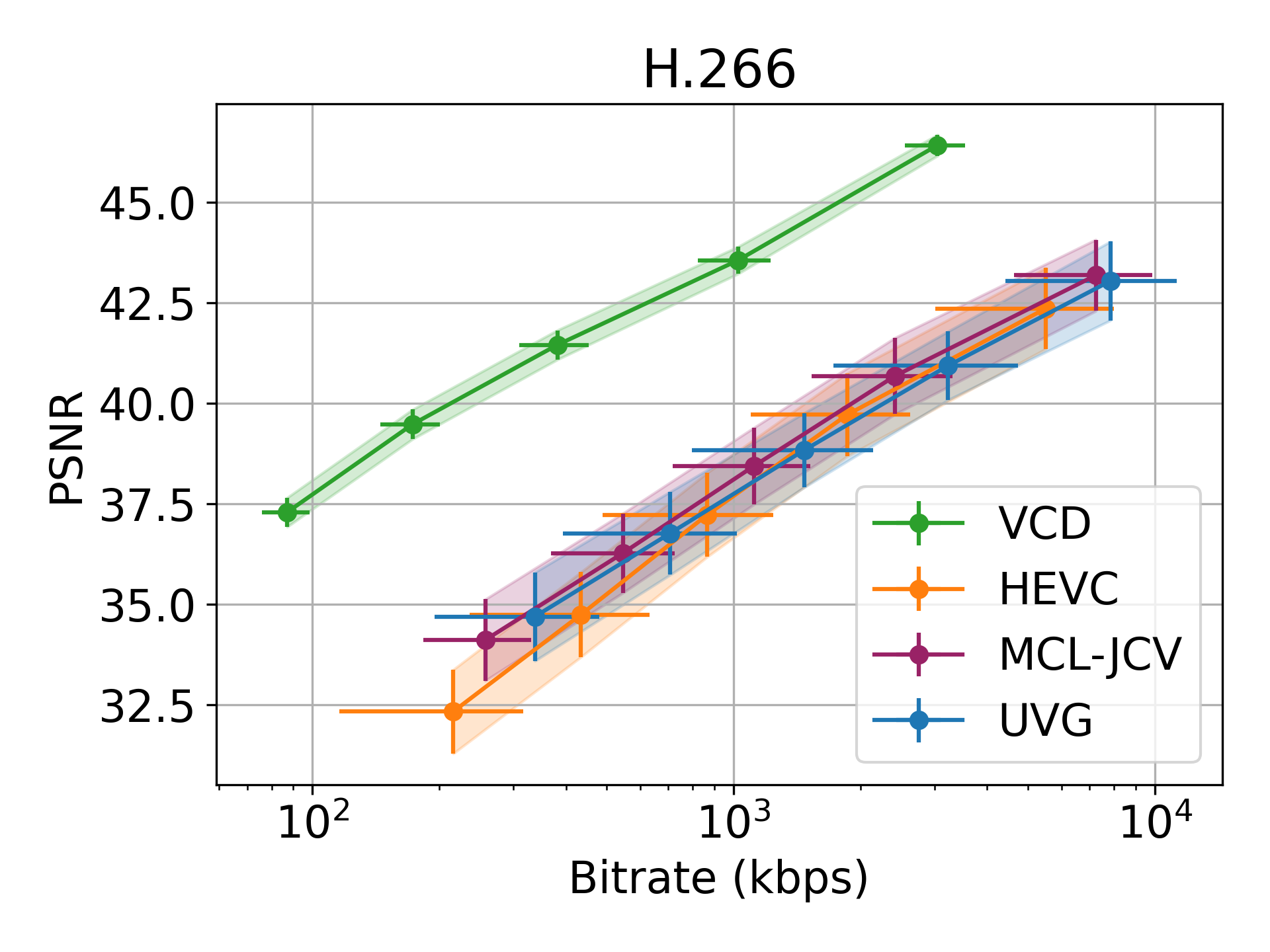}
        \caption{}
        \label{fig:vvc_psnr}
    \end{subfigure}
    \caption{Rate-distortion plots. (a-b) R-D plot of H.264, H.265, H.266, and AV1 on VCD dataset. (c-d) R-D plot of H.266 on all datasets. Error bars are 95\% confidence intervals for both distortion metric and bitrate.} 
    \label{fig:rd_plots}
\end{figure*}

\subsection{Recordings and selection procedures}
In our data collection process, video sequences were obtained through a crowd-sourcing platform. Participants were instructed to enact specific behaviors in front of their cameras as they were recorded to enact common video conferencing scenarios. We captured considerably more videos than present in VCD. This surplus enabled us to curate the videos to approximately match the desired quality and complexity distribution within each scenario. We only used video sequences with a resolution of 1080p or higher with at least 10 seconds duration and a person visible in the entire video duration. To produce a well-balanced coverage across diverse video qualities and complexity for this specific domain, we performed the following curation procedure for each scenario: Initially, we evaluated the subjective video quality of the captured video clips following the ITU-T Rec.~P.910 ACR test~\cite{itu-t_recommendation_p910_subjective_2021} and its crowdsourcing implementation~\cite{naderi_crowdsourcing_2022} and computed average SI and TI of each sequence. Subsequently, we categorized the videos into four quality brackets ([1,2.5), [2.5,3.5), [3.5, 4.5), [4.5,5)), and four SI/TI regions given the distribution of the entire set.
Finally, we selectively pruned the number of videos in each bracket to achieve an approximate distribution ratio of 20\%, 30\%, 30\%, and 20\% across the quality brackets and 25\% for each SI/TI region. During selection, the suitability of video sequences for subjective quality assessment (e.g., presence of edges, landmarks, and details), and the difficulty of sequence due to different factors like moving objects, lighting conditions, and contrast are considered. Finally, video clips in TH-OB and TH-BB are processed with the corresponding video pipelines and the video quality of the entire video set was evaluated in a separate subjective test. Figure \ref{fig:quality_distribution} shows the resulting distribution video quality of sequences within each scenario in the final set. As shown in Figure \ref{fig:thob}-\ref{fig:thbb}, background processing led to a different distribution of SI/TI than non-background processing.

\section{Analysis}
\label{sec:analysis}
In this section, we present the analysis of the effects of video quality and dataset composition on comparing video encoding performance of different codecs. We measured the performance of the codecs using the Bjøntegaard-delta rate (BD-Rate) metric~\cite{bjontegaard2001calculation}, which represents the percentage of saved bitrate compared to a baseline codec when delivering the same video quality~\cite{STP-VID-WPOM}. We report the average of the BD-Rates of the individual videos in each segment and 95\% confidence interval. We encoded the videos with four codecs, namely H.264 (baseline codec for calculating BD-Rate), H.265, H.266, and AV1.

\textbf{H.264 - H.265}: We encoded the videos using the Intel Quick Sync Video\cite{jiang2011intel} hardware encoder at five different quantization parameters, ranging from 20 to 44. We used a Surface laptop and applied a low-delay configuration (i.e., fast preset, no look ahead, large GOP, only one I-Frame).
For \textbf{H.266}, we used VVenC\cite{wieckowski2021vvenc} which provides a fast implementation of H.266. We encode videos using the low-delay and fast preset setting with no look ahead and large GOP size. We encoded each video with five quantization parameters from 22 to 42.
Similarly, we used 7 quantization parameters for \textbf{AV1} from 22 to 52.

\subsection{Sequence quality}
To illustrate the effect of video quality on codec performance, we divided VCD into three quality categories, namely high-quality ($MOS \geq 3.8$), medium-quality ($2.7\leq MOS < 3.8$), and low-quality ($MOS<2.7$), and evaluated the BD-Rates of H.265, H.266, and AV1 for each category. Table \ref{table:bd_rate_vs_quality} presents the average BD-Rates achieved by the codecs relative to H.264 on each video quality category. 

The result of two-way mixed ANOVA shows a significant main effect of the encoder ($F(2,314)= 1436$, $p<0.001$, $\eta^2=0.9$) and a significant main effect of source sequence quality ($F(2,157)= 9.17$, $p<0.001$, $\eta^2=0.11$) on the achieved BD-Rate over VMAF. A post hoc test using Holm-Bonferroni adjustment reveals that the codec's compression efficiency when applied on low and medium-quality sequences is significantly lower than their performance on high-quality source sequences, ($p=0.011$ and $p<0.001$, respectively). 
Similar results were observed when using PSNR instead of VMAF. 

\subsection{Scenarios}

Table~\ref{table:vcd-scenario} reports the compression efficiency of codecs on the VCD dataset when segmented into different scenarios (see Section~\ref{sec:dataset}). A two-way mixed ANOVA reveals that there is a significant interaction effect between the encoder and dataset scenarios on compression efficiency in terms of VMAF's BD-Rate ($F(6,312)= 23.02$, $p<0.001$, $\eta^2=0.31$). A similar interaction effect is also observed for PSNR's BD-Rate. The presence of these interaction effects in the model indicates that the relative performance of encoders on a given scenario may not replicate for other scenarios. 
Results of the post hoc test show that the compression efficiency in the scenario with background replacement (TH-OB) is significantly different from the other scenarios when measured over VMAF ($p<0.018$ for all combinations after Holm-Bonferroni correction). In addition, H.265's and H.266's efficiency are significantly different in mobile (TH-M) and desktop (TH) scenarios. H.265 achieved a much higher compression on mobile sequences compared to the desktop's (Hedges's $g=0.827$, $p= 0.004$). In contrast, H.266 performed considerably better on desktop scenarios (TH) compared to the mobile sequences (Hedges's $g=-0.856$, $p = 0.003$). A post hoc test also shows a significantly different BD-Rate calculated over PSNR in both scenarios where backgrounds are processed (TH-OB and TH-BB).

\subsection{Comparison with other datasets}
Table~\ref{table:bd_rate_vs_dataset} reports the compression efficiency of the codecs on the entirety of VCD, MCL-JCV, HEVC, and UVC datasets. One-way independent ANOVA models (one per codec) with Dataset as the independent factor and BD-Rate as the dependent variable show statistically significant differences between these datasets in terms of the codec performance measurement ($\eta^2_{H.266}=0.04 $, $\eta^2_{H.265}=0.053$, $\eta^2_{AV1}=0.044$). This difference indicates the challenges in predicting the performance of codecs on a specific dataset when it is optimized on general datasets. 
For further insights, we used a two-way mixed ANOVA model with both codec and dataset as independent variables. This model shows a significant interaction effect between the codec and dataset ($F(6,422) = 2.773$, $p =0.012$, $\eta^2=0.04$). The interaction effect indicates the potential for different patterns of BD-Rate by codec for each dataset. For example, on the HEVC dataset, the order of H.265 and AV1 is the reverse of the same codecs on the VCD dataset. 
Finally, Figures~\ref{fig:vcd_vmaf}-\ref{fig:vcd_psnr} show the rate-distortion curves of video codecs on the VCD dataset for PSNR and VMAF. Figure ~\ref{fig:vvc_vmaf}-\ref{fig:vvc_psnr} illustrates the performance of H.266 on all datasets, which shows a large difference in performance between VCD and general datasets.

\begin{table}[t]
\caption{Average (and 95\% CI) BD-Rates on VCD videos in different video quality categories.}
\label{table:bd_rate_vs_quality} 
\vspace{-0.5cm}
\begin{center}
\resizebox{\columnwidth}{!}{
    \begin{tabular}{l c c c c c c c}
    \toprule
    \textbf{Quality} & \textbf{N.} & \multicolumn{3}{c}{\textbf{BD-Rate on VMAF}}& \multicolumn{3}{c}{\textbf{BD-Rate on PSNR}}   \\    
    \textbf{segments}&\textbf{Clips} & H.265 & AV1 & H.266 & H.265 & AV1 & H.266 \\
    \midrule
    VCD - High &  61    & 38.1 (1.5) & 39.4 (2.0) & 67.9 (1.6) &  38.3 (1.5) & 41.2 (2.8) & 69.0 (1.8) \\   
    VCD - Medium & 70   & 34.6 (1.2) & 36.0 (1.9) & 64.1 (1.5) &  36.0 (1.4) & 38.2 (2.5) & 65.3 (1.9) \\
    VCD - Low &   29    & 33.1 (2.3) & 36.9 (3.7) & 63.2 (2.3) &  36.7 (2.4) & 41.4 (4.4) & 67.1 (2.2) \\
    
    \bottomrule
    
    \end{tabular}
}
\end{center}
\vspace{-0.3cm}
\end{table}

\begin{table}[t]
\caption{Average (and 95\% CI) BD-Rates on VCD videos in different scenarios.}
\label{table:vcd-scenario} 
\vspace{-0.5cm}
\begin{center}
\resizebox{\columnwidth}{!}{
    \begin{tabular}{l c c c c c c c}
    \toprule
    \textbf{Scenario} & \textbf{N.} & \multicolumn{3}{c}{\textbf{BD-Rate on VMAF}}& \multicolumn{3}{c}{\textbf{BD-Rate on PSNR}}   \\    
    \textbf{}&\textbf{Clips} & H.265 & AV1 & H.266 & H.265 & AV1 & H.266 \\
    \midrule
    TH      & 40 &  34.4 (2.0)  & 39.2 (2.5) & 68.0 (2.2) & 34.3 (2.1) & 39.9 (3.6) & 67.5 (2.5)\\   
    TH-OB   & 40 &  32.1 (1.5)  & 31.3 (2.3) & 67.1 (1.1) & 33.2 (1.5) & 30.4 (2.9) & 67.2 (1.6)\\
    TH-BB   & 40 &  37.1 (2.0)  & 38.8 (2.9) & 65.2 (1.6) & 42.1 (1.3) & 48.1 (2.1) & 70.8 (1.2)\\
    TH-M    & 40 &  38.9 (1.3)  & 40.5 (1.9) & 61.4 (2.5) & 38.4 (1.5) & 41.4 (2.6) & 62.6 (2.9)\\
    \bottomrule
    
    \end{tabular}
}
\end{center}
\end{table}

\begin{table}[t]
\caption{Average (and 95\% CI) BD-Rate of codecs in different datasets.}
\label{table:bd_rate_vs_dataset} 
\vspace{-0.5cm}
\begin{center}
\resizebox{\columnwidth}{!}{
    \begin{tabular}{l c c c c c c c}
    \toprule
    \textbf{Dataset} & \textbf{N.} & \multicolumn{3}{c}{\textbf{BD-Rate on VMAF}}& \multicolumn{3}{c}{\textbf{BD-Rate on PSNR}}   \\    
    \textbf{}&\textbf{Clips} & H.265 & AV1 & H.266 & H.265 & AV1 & H.266 \\
    \midrule
    VCD (ours)& 160& 35.7 (0.9) & 37.5 (1.3) & 65.4 (1.0) & 37.0 (1.0) & 39.9 (1.7) & 67.0 (1.2)\\
    UVG & 16 &41.1 (7.4) & 44.9 (7.0) & 70.0 (6.6) & 42.9 (6.4) & 44.3 (6.6) & 73.6 (6.6)\\
    MCL-JCV & 30& 39.1 (3.0) & 39.9 (4.6) & 63.4 (3.4) & 39.9 (3.6) & 37.7 (5.4) & 66.8 (3.6)\\
    HEVC\textsuperscript{*} & 9&  38.2 (6.0) & 36.6 (3.3)& 62.1 (7.3) & 35.6 (6.1) & 32.1 (5.9) & 64.7 (7.3)\\
    \bottomrule
    \multicolumn{8}{l}{\textsuperscript{*} Only 1080 clips to be consistent with other datasets.}
    \end{tabular}
}
\end{center}
\vspace{-0.3cm}
\end{table}

\section{Conclusions}
\label{sec:conclusions}

We have created and open-sourced the VCD dataset for the video conferencing use case which includes 160 unique subjects and environments in four popular video conferencing scenarios. We have shown that four common video codecs perform statistically differently with it than general purpose datasets. 
These findings highlight the importance of developing and benchmarking video conferencing compression solutions using a domain-specific dataset. Our work underscores the significance of VCD as a valuable resource for researchers and practitioners in this field, enabling them to refine and optimize video codecs for enhanced video conferencing experiences.
The fact that video sequences are recorded by a large set of user devices that are typically employed in video conferencing sessions increases the ecological validity of models benchmarked on this dataset.
The versatility of VCD can be increased by including additional pertinent video scenarios, such as multi-headed conference room videos, whiteboard videos, lecture or stage videos, screen sharing, and even animated avatar videos. In future work, we will evaluate the performance of end-to-end ML-codecs that perform video enhancement as well on this dataset.

\newpage
\bibliographystyle{IEEEbib}
\bibliography{IC3-AI, other_ref}
\end{document}